\documentclass{rspublic}

\usepackage{graphicx}

\begin{document}
\renewcommand{\r}{\mathbf{r}}
\newcommand{\ha}{\hat{a}}
\newcommand{\hr}{\hat{r}}
\newcommand{\ren}{\mathrm{ren}}
\newcommand{\lin}{\mathrm{lin}}
\newcommand{\Manning}{\mathrm{Manning}}

\newlength{\GraphicsWidth}
\setlength{\GraphicsWidth}{10cm}

\title{Nonlinear screening of charged macromolecules}

\author{Gabriel T\'ellez}
\affiliation{Departamento de F\'{\i}sica, Universidad de los Andes,
  A.A. 4976, Bogot\'a, Colombia}

\maketitle

\begin{abstract}{Colloids, electrolytes, nonlinear Poisson--Boltzmann equation}
We present several aspects of the screening of charged macromolecules
in an electrolyte. After a review of the basic mean field approach,
based on the linear Debye--H\"uckel theory, we consider the case of
highly charged macromolecules, where the linear approximation breaks
down and the system is described by full nonlinear Poisson--Boltzmann
equation. Some analytical results for this nonlinear equation give
some interesting insight on physical phenomena like the charge
renormalization and the Manning counterion condensation.
\end{abstract}

\section{Introduction and the linear Poisson--Boltzmann equation}

A colloidal suspension is a system composed of two substances, one
dispersed into the other. The dispersed phase is composed of
macromolecules, with size of the order $10^{-8}$ m -- $10^{-6}$ m,
while the dispersion medium, or continuous medium, is composed of
small micromolecules, and/or microions, with size of the order of the
nanometer. Of particular interest are the charge-stabilized colloids,
where the dispersed macromolecules have ionizable sites and, when
immersed into the dispersion medium, they acquire a surface electric
charge which ensures repulsion between them and thus allows the
colloid to stabilize and prevents aggregation.

Since the length and time scales of the microcomponents of the
dispersion medium are much smaller than the ones of the dispersed
phase, it is convenient to average over the degrees of freedom of the
dispersion medium and treat the system as a one-component system
composed by the macromolecules which interact via an effective
potential. To understand the physical and thermodynamical properties
of these systems, it is important to determine this effective
interaction between the macromolecules, which results not only from
the direct interaction between the macromolecules, but also the
interaction mediated by the microions of the dispersion medium.

The basic theory to find this effective interaction for
charge-stabilized colloid was developed independently by~\cite{DL},
and~\cite{VO}, and it is known as the DLVO theory. It is based on the
work of~\cite{DH}. Let us consider a spherical charged macromolecule,
with radius $a$, and charge $Ze$ ($e$ is the elementary charge)
immersed in an electrolyte with positive microions of charge $z_{+} e$
and average density $n_{+}$, and negative microions of charge $-z_{-}
e$ and average density $n_{-}$. Without loss of generality we can
suppose $Z>0$. \cite{DH} idea is to treat the microions in a mean
field approximation: at temperature $T$, the local density of
microions of charge $q_{\pm}=\pm z_{\pm} e$ at a distance $r$ from the
macromolecule can be approximated by
\begin{equation}
  n_{\pm}(r)=n_{\pm} e^{-\beta q_{\pm} \Psi(r)} \,,
\end{equation}
where $\Psi(r)$ is the electrostatic potential, and $\beta =1/(k_B
T)$, with $k_B$ the Boltzmann constant. Replacing this into Poisson
equation of electrostatics yield the Poisson--Boltzmann equation
\begin{equation}
  \label{eq:PBcomp}
  \Delta \Psi = -\frac{4\pi e}{\epsilon} \left( z_{+} n_{+} e^{-\beta
    e z_{+} \Psi} - z_{-} n_{-} e^{\beta e z_{-} \Psi} \right) \,,
\end{equation}
where $\epsilon$ is the dielectric constant of the dispersion medium.

It is convenient to introduce the following notations: the reduced
potential $y=\beta e \Psi$, the Bjerrum length $l_B=\beta
e^{2}/\epsilon$, the Debye length $\kappa^{-1}=(4\pi l_B (z_{+}^2
n_{+} + z_{-}^2 n_{-}))^{-1/2}$. With these notations,
Poisson--Boltzmann equation reads
\begin{equation}
  \label{eq:PB}
  \Delta y = \frac{\kappa^2}{z_{+}+z_{-}}\left[ e^{z_{-} y} -
    e^{-z_{+} y} \right]
  \,.
\end{equation}
If the electrostatic coupling between the macromolecule and the
microions is small, $y(r)\ll 1$, for any distance $r$, the nonlinear
Poisson--Boltzmann equation~(\ref{eq:PB}) can be linearized to obtain
\begin{equation}
  \label{eq:DH}
  \Delta y = \kappa^{2} y
  \,.
\end{equation}
For an impenetrable spherical macromolecule with uniform surface
charge (total charge $Ze$ and radius $a$), the solution of this equation
is the DLVO potential
\begin{equation}
  \label{eq:DLVO}
  y(r)=Z l_B \frac{e^{\kappa a}}{1+\kappa a} \frac{e^{-\kappa r}}{r}
  \,.
\end{equation}
From this equation, one can see that $1/\kappa$ is the screening
length.

It is also interesting to consider the case of cylindrical
macromolecules, for instance stiff polyelectrolytes, ADN, etc. As a
first approximation, for an infinitely long cylinder with radius $a$
and linear charge density $e/\ell$, uniformly spread over its surface,
the solution of equation~(\ref{eq:DH}) gives the electrostatic
potential at a radial distance $r$ from the cylinder
\begin{equation}
  \label{eq:ylin}
  y(r)=\frac{2\xi}{\ha K_1(\ha)}\,K_0(\hr)
  \,,
\end{equation}
where $K_0$ and $K_1$ are the modified Bessel functions of order 0 and
1. We have defined the reduced linear charge density of the cylinder
$\xi=l_B/\ell$, and it is convenient to measure the distances in Debye
length units: $\hr=\kappa r$ and $\ha = \kappa a$. The boundary
conditions that complement the differential equation~(\ref{eq:DH}) to
yield the solution~(\ref{eq:ylin}) are
\begin{equation}
  \label{eq:BC}
  \lim_{r\to a} \,r\, \frac{dy}{dr} \,=\, -2 \,\xi
  \,,\quad\text{and}\quad \lim_{r\to\infty} \nabla y(r)=0
  \,.
\end{equation}
It should be noticed that at large distances from the cylinder,
compared to the Debye length, the potential exhibits again an
exponential decay, as for the case of spherical macromolecules,
\begin{equation}
  y(r)\sim \frac{2 \xi}{\ha K_1(\ha)} \sqrt{\frac{\pi}{2\kappa r}} \,
  e^{-\kappa r}\,,\qquad r\gg \kappa^{-1}
  \,.
\end{equation}
In the following sections, we turn our attention to the case of highly
charged macromolecules, where the linear approximation breaks down and
the full nonlinear equation~(\ref{eq:PB}) should be used. Several
nonlinear effects appear, which we will review. In
section~\ref{sec:charge-renorm}, we study the nonlinear phenomenon
known as the charge renormalization~\cite{Alexander, TrizacPRL}, which
is generic, both for the spherical and cylindrical geometries. In
section~\ref{sec:cyl}, we focus our attention on the cylindrical case,
where an analytical solution of the nonlinear Poisson--Boltzmann
equation is available. From the analysis of this analytic solution we
discuss the phenomenon of counterion condensation.

\section{Charge renormalization}
\label{sec:charge-renorm}

For a highly charged macromolecule, Poisson--Boltzmann
equation~(\ref{eq:PB}) cannot be linearized near the macromolecule
surface. However, due to the screening effect, the potential will
decay and become small, $|y(r)|\ll1$, at large distances from the
macromolecule surface, $r-a \gg \kappa^{-1}$. In that far region, the
linear version~(\ref{eq:DH}) of Poisson--Boltzmann equation
holds. Then, for a spherical macromolecule, the potential will behave
as
\begin{equation}
  \label{eq:sollin}
  y(r)\sim A \frac{e^{-\kappa r}}{r}
\end{equation}
at large distances from the macromolecule. To find the constant of
integration $A$, one needs to enforce the boundary condition at the
surface of the macromolecule that the normal component of the electric
field is proportional to the surface charge density. However, the
form~(\ref{eq:sollin}) of the potential is not valid in that close
region. One needs to find also the form of the potential close to the
macromolecule surface, and connect it to the large-distance
behavior~(\ref{eq:sollin}) to find explicitly the integration constant
$A$. In analogy to the linear solution~(\ref{eq:DLVO}), one can write
the $A=Z_{\ren}l_B e^{\kappa a}/(1+\kappa a)$, defining a renormalized
charge $Z_{\ren}$. The large-distance behavior of the potential then
takes a DLVO familiar form
\begin{equation}
  \label{eq:yZren}
  y(r)\sim  Z_{\ren} l_B\,
  \frac{e^{\kappa a}}{1+\kappa a}\frac{e^{-\kappa r}}{r}
  \,,
\end{equation}
but replacing the bare charge $Z$ of the macromolecule by the
renormalized one $Z_{\ren}$. In the cylindrical geometry, the
renormalized charge concept also applies. In that case, the
large-distance behavior of the potential is
\begin{equation}
  \label{eq:Xiren}
  y(r)\sim\frac{2\xi_{\ren}}{\ha K_1(\ha)}\,K_0(\hr)
    \,,
\end{equation}
with a renormalized linear charge density $\xi_{\ren}$. The
determination of the renormalized charge requires knowledge of the
short-distance behavior of the solution of the nonlinear
Poisson--Boltzmann equation~(\ref{eq:PB}). This can be done
numerically, as in the original work of~\cite{Alexander}. In
experimental situations, the unknown renormalized charge is often
taken as an adjustable fitting parameter. There are also analytical
approaches to find the renormalized charge~\cite{TrizacPRL, Shkel,
  Tellez21}, mostly based on approximations using the solution to the
nonlinear Poisson--Boltzmann equation in the planar case.

Let us illustrate the concept of charge renormalization in the planar
geometry where an explicit solution for the nonlinear
Poisson--Boltzmann equation is known~\cite{Gouy, Chapman}. The system
is an infinite charged plane, with charge density $\sigma>0$, immersed
in an electrolyte, which, for simplicity, we consider symmetric
$z_{+}=z_{-}=1$. Let $Ox$ be the axis perpendicular to the plane,
which we suppose located at $x=0$, the electrolyte occupies the region
$x>0$. The nonlinear Poisson--Boltzmann equation in this situation
reads
\begin{equation}
  \frac{d^2 y(x)}{dx}=\kappa^2 \sinh y(x)
  \,.
\end{equation}
It can be integrated once by multiplying by $dy/dx$,
\begin{equation}
  \frac{dy(x)}{dx}=-2\kappa\sinh\frac{ y(x)}{2}
  \,.
\end{equation}
where the boundary condition $dy/dx\to0$ when $x\to\infty$ has been
used. This last equation is separable and can be integrated, finally
obtaining
\begin{equation}
  \label{eq:PB1dsol}
  y(x)=2\ln \frac{1+Ae^{-\kappa x}}{1-Ae^{-\kappa x}}
  \,.
\end{equation}
Where $A$ is a constant of integration, which is found using the
boundary condition at the surface of the charged plane
\begin{equation}
  \label{eq:BC1d}
  \frac{dy}{dx}(0)=-4\pi l_B \sigma/e
  \,.
\end{equation}
Notice that at large distances
from the plane, $\kappa x\gg 1$, the potential behaves as
\begin{equation}
  \label{eq:ynonlin1d}
  y(x)\sim 4A e^{-\kappa x}
  \,.
\end{equation}
This is the expected behavior for the linear version of
Poisson--Boltzmann equation in this geometry $y_{\lin}''(x)-\kappa^2
y_{\lin}(x)=0$. The linear solution is $y_{\lin}(x)= 4\pi l_B
\sigma e^{-\kappa x} /(e\kappa)$. Comparing to the large-distance
behavior of the nonlinear solution~(\ref{eq:ynonlin1d}), one can
define the renormalized surface charge density $\sigma_{\ren}$, by
writing the constant of integration $A$ as 
\begin{equation}
  \label{eq:A1d-sigmaren}
A=\pi l_B
\sigma_{\ren}/(e\kappa)
\,.
\end{equation}
Then, the large-distance behavior of the
potential is
\begin{equation}
  y(x)\sim \frac{4\pi l_B \sigma_{\ren}}{e\kappa} \, e^{-\kappa x}
  \,.
\end{equation}
To find explicitly the renormalized charge, one needs to apply the
boundary condition~(\ref{eq:BC1d}) at the surface of the charged
plane. For this, one needs the short-distance behavior of the
potential. From the explicit solution~(\ref{eq:PB1dsol}), we find
\begin{equation}
  y(x)=2 \ln\frac{1+A}{1-A} - \frac{4A x}{1-A^2}+O(x^2)\,,
\end{equation}
when $x\to 0$. Using this, we apply the boundary
condition~(\ref{eq:BC1d}) to find
\begin{equation}
  \frac{A}{1-A^2}=\frac{\pi \sigma l_B}{e\kappa}
  \,.
\end{equation}
Solving, and using~(\ref{eq:A1d-sigmaren}), we find the renormalized
surface charge
\begin{equation}
  \hat{\sigma}_{\ren} = \frac{\sqrt{\hat{\sigma}^2+1}-1}{\hat{\sigma}}
  \,,
\end{equation}
where we have defined reduced charge densities $\hat{\sigma}=\pi l_B
\sigma/(e\kappa)$ and $\hat{\sigma}_{\ren}=\pi l_B
\sigma_{\ren}/(e\kappa)$.  

%%%%
\begin{figure}
  \begin{center}
      \includegraphics[width=\GraphicsWidth]{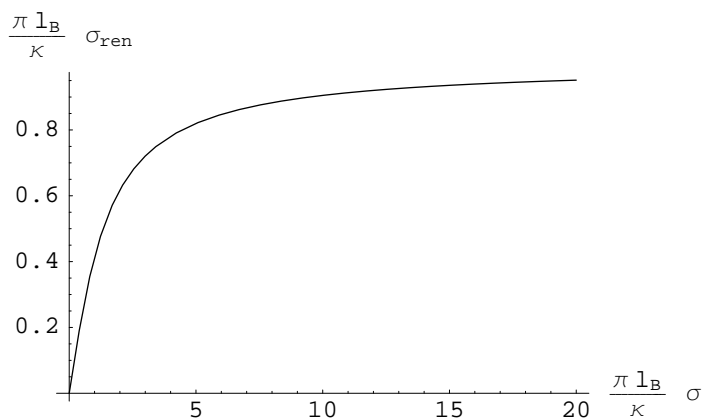}
  \end{center}
  \caption{Renormalized charge as a function of the bare charge in the
    planar case, for a symmetric electrolyte.
    \label{fig:sigmaren-plan}}
  
\end{figure}

%%%%

Figure~\ref{fig:sigmaren-plan} shows a plot of the renormalized charge
as a function of the bare charge. Notice the saturation effect: when
$\sigma \to \infty$, the renormalized charge approaches a finite value
$\hat{\sigma}_{\ren}\to 1$. This saturation effect can also appear in
other theories obtained by modification of Poisson--Boltzmann equation
using a density functional formulation~\cite{Tellez-sat}. In that
saturation regime, the large-distance behavior potential becomes
independent of the bare charge of the plane
\begin{equation}
  y_{\mathrm{sat}}(x) \sim y_0 e^{-\kappa x} = 4  e^{-\kappa x}
  \,.
\end{equation}
The value $y_0=4$ plays an important role. It can be seen as an
effective surface potential for the plane, if one wants to match the
linear solution of Poisson--Boltzmann equation with the nonlinear one
in the close vicinity of the highly charged plane. This is also the
starting point to find the renormalized charge at saturation for
highly charged macromolecules of arbitrary shape. For these
macromolecules, when they are highly charged, the linear
Poisson--Boltzmann equation can be solved with an effective boundary
condition of constant surface potential $y_0$, to find the behavior of
the potential at large distances and the corresponding renormalized
charge, as explained in~\cite{Bocquet-JCP}. For instance, for a
spherical macromolecule, the solution of the linear Poisson--Boltzmann
equation with the effective constant potential boundary condition at
the surface of the macromolecule $y(a)=y_0=4$, is
\begin{equation}
  \label{eq:ysat-sphere}
  y_{\text{sat, sphere}}(r)= y_0 e^{\kappa a} a\, \frac{e^{-\kappa
      r}}{r}
  \,.
\end{equation}
The nonlinear solution, at large distances has the behavior given by
equation~(\ref{eq:yZren}). Comparing both equations~(\ref{eq:yZren})
and~(\ref{eq:ysat-sphere}), we find an approximate value for the
renormalized charge in the saturation regime
\begin{equation}
  Z_{\ren}^{\text{sat}} = \frac{a}{l_B} y_0 (\kappa a + 1) =
  \frac{a}{l_B} (4 \kappa a + 4) 
  \,.
\end{equation}
This approximation is based on the planar solution of the nonlinear
Poisson--Boltzmann equation, and therefore it is accurate for large
macromolecules with $\kappa a\gg 1$, and only at the first order in
$\kappa a$. One can improve this estimate by developing a planar
expansion of the solution of the spherical geometry, as done
in~\cite{Shkel, Trizac-analytical, Tellez21}. Up to terms of order
$O(1/(\kappa a))$, the renormalized charge at saturation for spheres
in a 1:1 electrolyte is~\cite{Trizac-analytical}
\begin{equation}
  Z_{\ren}^{\text{sat}} = 
  \frac{a}{l_B} (4 \kappa a + 6) 
  \,.
\end{equation}

The starting point to obtain estimates of the renormalized charge at
saturation is the value $y_0$ of the effective surface potential at
saturation in the planar geometry. This depends only on the
constitution of the electrolyte. We propose now a simple formula that
gives $y_0$ in the generic case of a multicomponent electrolyte,
composed of several species of ions with charges $\{q_{\alpha} e\}$
and densities $\{n_{\alpha}\}$. The nonlinear Poisson--Boltzmann
equation in the planar geometry reads now
\begin{equation}
  y''(x)+ 4\pi l_B \sum_{\alpha} q_{\alpha} n_{\alpha}
  e^{-q_{\alpha} y(x)}=0\,.
\end{equation}
Multiplying by $y'(x)$ this equation, it can be integrated once to
find
\begin{equation}
  (y'(x))^{2}= 8 \pi l_B \sum_{\alpha} n_{\alpha} (e^{-q_{\alpha}
    y(x)}-1)
  \,,
\end{equation}
where the boundary condition $y'(x)\to0$ when $x\to\infty$ has been
used.  Introducing the inverse Debye length $\kappa=(4\pi l_B
\sum_{\alpha} q_{\alpha}^{2} n_{\alpha})^{-1/2}$, one obtains the
formal solution
\begin{equation}
  \label{eq:formal1d}
  \kappa x = \int_{y(x)}^{y(0)} \frac{du}{\sqrt{
      \frac{2}{\sum_{\alpha} q_{\alpha}^{2} n_{\alpha}}\sum_{\alpha}
      n_{\alpha}(e^{-q_{\alpha} u}-1)}}
    \,.
\end{equation}
Suppose the charged plane is located at $x=0$ and positively charged,
and we are in the saturation regime, therefore $y(0)\to+\infty$. At
large distances from the plane, $\kappa x\gg 1$, the potential behaves
as $y(x)\sim y_0 e^{-\kappa x}$, thus $\kappa x = \ln y_0 - \ln y(x) +
o(\ln y(x))$ as $y(x)\to 0$. Replacing in~(\ref{eq:formal1d}) we find
\begin{eqnarray}
  \label{eq:y0prelim}
\ln y_0 &=& \lim_{y\to 0} \left[
\int_{y}^{\infty}
\frac{du}{\sqrt{
      \frac{2}{\sum_{\alpha} q_{\alpha}^{2} n_{\alpha}}\sum_{\alpha}
      n_{\alpha}(e^{-q_{\alpha} u}-1)}}
+ \ln y
\right]
\\
&=&
\text{Pf.} 
\int_{0}^{\infty}
\frac{du}{\sqrt{
      \frac{2}{\sum_{\alpha} q_{\alpha}^{2} n_{\alpha}}\sum_{\alpha}
      n_{\alpha}(e^{-q_{\alpha} u}-1)}}
\,.  
\nonumber
\end{eqnarray}
Thus, the value of $y_0$ is expressed as a Hadamard finite part (Pf.)
of the integral~(\ref{eq:y0prelim}). Alternatively, it can be computed
from
\begin{equation}
\ln y_0 = 
\int_{0}^{1}
\left[
\frac{1}{\sqrt{
      \frac{2}{\sum_{\alpha} q_{\alpha}^{2} n_{\alpha}}\sum_{\alpha}
      n_{\alpha}(e^{-q_{\alpha} u}-1)}}
-\frac{1}{u}\right]du
+
\int_{1}^{\infty}
\frac{du}{\sqrt{
      \frac{2}{\sum_{\alpha} q_{\alpha}^{2} n_{\alpha}}\sum_{\alpha}
      n_{\alpha}(e^{-q_{\alpha} u}-1)}}
\,.
\end{equation}

In the case of a two-component electrolyte $q_{1}= z_{+}$ and
$q_{2}=-z_{-}$, the saturation value $y_0$ can be expressed as a
function of the ratio $r=z_{+}/z_{-}$,
\begin{equation}
  y_0=\frac{1}{z_{-}} \exp\left[
    \text{Pf.}\int_{0}^{\infty} \sqrt{\frac{r}{2}}\frac{du}{\sqrt{
        \frac{e^{-r u}}{1+r}+
        \frac{e^{u}}{1+r^{-1}}-1}}
\right]\,.
\end{equation}
In the cases of electrolytes with $z_{+}:z_{-}$ equal to 1:1, 1:2, and
2:1, the integral can be computed exactly to find the known
values~\cite{Tellez21}, $y_0^{1:1}=4$,
$y_0^{1:2}=6(2-\sqrt{3})\simeq 1.6077$, and
$y_0^{2:1}=6$. Table~\ref{tab:ysat} gives the value of the saturation
potential for other electrolytes.
\begin{table}
  \caption{\label{tab:ysat}Value of the saturation potential for several
electrolytes.}
  \begin{tabular}{|c||c|c|c|c|c|c|c|c|c|}
    \hline
    $z_+:z_-$  &  1:5 &  1:4 & 1:3    &   1:2    &     1:1       &
    2:1      &    3:1   &  4:1 & 5:1     \\
    \hline
    $y_0$ &  0.56080 & 0.71742  & 0.99388 & 1.6077 & 4 & 6 &
    8.7070 & 12.314 & 17.337 \\
    \hline
    \\
    \hline
    $z_+:z_-$ &   2:3 & 3:2 & 2:5 &  5:2&  & 3:5 & 5:3 & 1:10 & 10:1 \\
    \hline
    $y_0$ &  1.1542 & 2.4611 & 0.61471 & 3.6270 & & 0.67242 & 1.7544 &
    0.26761 & 70.337 \\
    \hline
  \end{tabular}
\vspace{2mm}
\end{table}

\section{Cylindrical macromolecules}
\label{sec:cyl}

\subsection{Exact solutions for Poisson--Boltzmann equation and the
  connection problem}

In this section, we focus our attention on the study of the screening
of a thin cylindrical macromolecule, with radius $a\ll \kappa^{-1}$.
Besides the charge renormalization effect, another interesting
phenomenon that occurs in this geometry is the counterion
condensation. This was first realized by Onsager and studied
by~\cite{Manning} and~\cite{Oosawa}. To understand this phenomenon,
consider the Boltzmann factor between the macromolecule and an ion of
opposite charge (counterion): $\exp(-2 z_{-} \xi \ln r)=r^{-2z_{-}
  \xi}$. It diverges when $r\to0$, and furthermore it is not
integrable near $r\to 0$ if $\xi > 1/z_{-}$. This means that, for an
infinitely thin macromolecule, with radius $a=0$, the thermodynamics
are not properly defined unless $\xi < 1/z_{-}$. In real situations
$a\neq 0$. For $\kappa a\ll 1$, when $\xi > 1/z_{-}$, the density of
counterions will be very large near the surface of the
macromolecule. These counterions are bounded to the macromolecule:
besides the diffuse screening cloud of ions around the macromolecule,
there is also a thin layer of condensed counterions very near to the
surface of the cylinder.

This counterion condensation effect can be studied quantitatively in
the mean field approximation, since an analytical solution of the
nonlinear Poisson--Boltzmann equation in the cylindrical geometry is
available. For a 1:1 electrolyte, this solution was found
by~\cite{McCoy}, in a different context, in relation to the
correlation functions of the two-dimensional Ising model. Later on,
\cite{Widom} developed the solution for the asymmetric cases 2:1 and
1:2, and \cite{TracyWidom} studied the short-distance asymptotics of
the solution and solved the problem of connecting the large-distance
and the short-distance behaviors of the solution. The consequences of
this mathematical work to the screening of cylindrical macromolecules
were reported by~\cite{McCaskill}, \cite{TracyWidom-poly} and
\cite{TrizacTellez-Manning}. A more extensive study is
presented in~\cite{TellezTrizac-PBcyl-exact}. We summarize here some
of the main findings of that work.

The short-distance behavior of the potential can be obtained by a
physical argument. Very close to the charged cylindrical
macromolecule, one would expect the potential to be the bare Coulomb
potential $-2A\ln r+\text{constant}$, with $A$ some constant related to
the charge of the macromolecule. By replacing this ansatz into
Poisson--Boltzmann equation~(\ref{eq:PB}) one can compute
systematically the following terms of the short-distance expansion, to
find~\cite{TracyWidom, TellezTrizac-PBcyl-exact,
  TrizacTellez-macromol}
\begin{equation}
y(r)=
-2A\ln\hr
+2\ln B
-2\ln\left[1-\frac{B^2\hr^{2-2A}}{16(1-A)^2}
    \right]+O(\hr^{2+2A})
\,.
\label{eq:y-short}
\end{equation}
We specialize here in the 1:1 electrolyte. For the general case of a
$z_{+}$:$z_{-}$ electrolyte, see~\cite{TrizacTellez-macromol}. $A$ and
$B$ are some constants of integration. The constant $A$ can be related
to the charge density of the electrolyte by writing the first
boundary condition~(\ref{eq:BC})
\begin{equation}
  \label{eq:xi-A}
  \xi = A - \frac{(2-2A) (\kappa a)^{2-2 A} B^{2}}{16 (1-A)^2
    -B^2(\kappa a)^{2-2 A} }
  \,.
\end{equation}
From the previous section, we already know the large-distance behavior
of the potential, it is the screened potential~(\ref{eq:Xiren})
\begin{equation}
  y(r) \sim 4\lambda K_0(\hr)\,,
\end{equation}
where $\lambda$ is some constant related to the renormalized charge by
\begin{equation}
  \label{eq:xiren-lambda}
  \xi_{\ren}=\ha K_{1}(\ha) \lambda/2
  \,.
\end{equation}
 By using the explicit analytical solution of the nonlinear
 Poisson--Boltzmann equation from~\cite{McCoy} and \cite{Widom},
 \cite{TracyWidom} where able to solve the connection problem of
 relating the constants of integration from the short-distance behavior
 $A$ and $B$, to the one of the large-distance behavior $\lambda$. To
 satisfy the boundary condition $y'(r)\to0$ when $r\to\infty$, the
 constants $A$ and $B$ need to satisfy
 \begin{equation}
   \label{eq:B}
   B=2^{3A}\frac{\Gamma\left(\frac{1+A}{2}\right)
   }{\Gamma\left(\frac{1-A}{2}\right)}
   \,,
 \end{equation}
where $\Gamma$ is the gamma function, and, $A$ and $\lambda$ need to
satisfy
\begin{equation}
  \label{eq:connect}
  \lambda=\frac{1}{\pi}\sin\left(\frac{\pi A}{2}\right)
  \,.
\end{equation}
The first physical consequence of these relations, is that we can
obtain an analytical expression for the renormalized charge, by
combining equations~(\ref{eq:xi-A}), (\ref{eq:xiren-lambda})
and~(\ref{eq:connect}). In the simplest situation, when $a=0$, this
gives, $A=\xi$ and
\begin{equation}
  \label{eq:xiren-exact}
  \xi_{\ren}=\frac{2}{\pi}\,\sin\frac{\pi\xi}{2}
  \,.
\end{equation}

\subsection{Counterion condensation}

The previous discussion, and in particular equation~(\ref{eq:y-short})
are only valid provided that $A<1$. Indeed, if $A=1$,
equation~(\ref{eq:y-short}) becomes singular: the last term becomes of
the same order as the second, and besides, the constant $B$ from
equation~(\ref{eq:B}) becomes undefined. This is the mathematical
signature of the counterion condensation phenomenon. Notice that for
$a=0$, the constant $A$ is the linear charge of cylinder $A=\xi$, and
the value $\xi_{\Manning}=1$ is precisely the threshold for counterion
condensation discussed earlier.

For a cylinder with nonzero radius $a\neq 0$, notice that, using
equation~(\ref{eq:xi-A}), the threshold $A=\xi_{\Manning}=1$ corresponds
for the linear charge to the threshold
\begin{equation}
  \label{eq:xic}
  \xi_{c}=1+\frac{1}{\ln\ha + C}
  \,,
\end{equation}
with $C=\gamma-3\ln 2\simeq-1.502$, with $\gamma$ the Euler
constant. Note that there is a negative logarithmic correction in the
radius $a$ of the cylinder to the Manning value $\xi_{\Manning}=1$ for
the threshold for condensation: $\xi_{c}\leq 1$.

To extend the solution beyond the condensation threshold,
\cite{TracyWidom} suggested to write $A$ as a complex number
$A=1-i\mu/2$. Replacing into equation~(\ref{eq:y-short}), the
short-distance expansion of the potential now reads
\begin{equation}
  \label{eq:yshort-condens}
  y(r)=  - 2 \ln \hr
  -2 \ln\frac{\sin(-2\mu \ln \hr - 2\mu C)}{4\mu}
  \,.
\end{equation}
The constant $\mu$ can be expressed in terms of the bare linear charge
density $\xi$ by using the first boundary condition~(\ref{eq:BC}), and
connected to the large-distance expansion of the potential and the
renormalized charge by means of equation~(\ref{eq:connect}), replacing
$A=1-i\mu/2$, for details see~\cite{TellezTrizac-PBcyl-exact}. 

\begin{figure}
  \begin{center}
      \includegraphics[width=\GraphicsWidth]{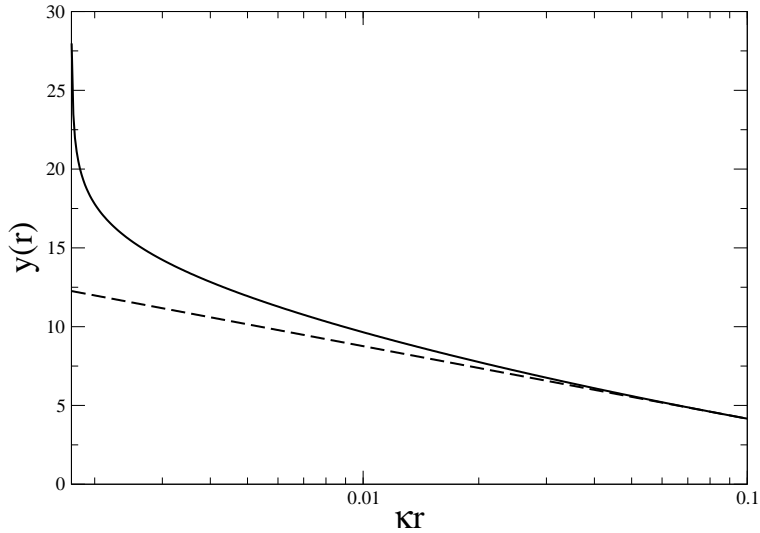}
  \end{center}
  \vspace{2mm}
  \caption{Short-distance expansion of the
    potential~(\ref{eq:yshort-condens}) (thick line) compared to
    $-2\ln(\hr/(4\mu))$ (dashed line). Parameter $\mu=0.2$. 
    \label{fig:y-short}}
  
\end{figure}

Notice the first term of expansion~(\ref{eq:yshort-condens})~: $-2\ln
\hr$. It is the bare Coulomb potential of a charged line with linear
charge density $\xi_{\Manning}=1$. This a characteristic of the
counterion condensation phenomenon. At intermediate distances of the
charged cylinder, one ``sees'' a cylinder with an effective charge
$\xi_{\Manning}=1$, if the bare charge exceeds the threshold value
$\xi_{c}$. The second term of equation~(\ref{eq:yshort-condens}) can
become very large in the close proximity of the cylinder. This second
term represents the thin layer of condensed counterions located at the
surface of the cylinder. Figure~\ref{fig:y-short}, shows a plot of the
potential close to the cylinder, and compares it to the bare term
$-2\ln \hr$.

%%%%

%%%%

\section{Concluding remarks}

To summarize and conclude the review presented here, we would like to
stress the differences between the different linear charges densities
that we presented for cylindrical macromolecules: renormalized charge,
Manning charge, threshold charge for counterion condensation.

The renormalized charge characterizes the behavior of the potential
far from the charged macromolecule. At those large distances, $r\gg
\kappa^{-1}$, the potential exhibits an exponential decay. The
prefactor of this exponential decay is proportional to the
renormalized charge, as shown in equation~(\ref{eq:Xiren}).

At short distances, there are two possible behaviors, depending on the
value of the bare linear charge density $\xi$ compared to the
threshold value $\xi_{c}$ given by equation~(\ref{eq:xic}). If
$\xi<\xi_{c}$, the potential behaves as given by
equation~(\ref{eq:y-short}). It is a bare Coulomb potential with a
prefactor given in terms the bare charge of the macromolecule. If
$\xi>\xi_{c}$, the counterion condensation takes place. The
short-distance behavior of the potential is given by
equation~(\ref{eq:yshort-condens}).  A thin layer of counterions is
bound to the surface of the cylinder, which makes the potential very
large in that region. Beyond this layer, the potential behaves as
$-2\xi_{\Manning}\ln r$~: the bare Coulomb potential for a charged
cylinder but with an effective charge $\xi_{\Manning}=1$, the Manning
charge for counterion condensation, see
figure~\ref{fig:y-short}. Notice that if the radius of the cylinder
$a\neq 0$, the Manning value differs from the threshold value:
$\xi_{\Manning} > \xi_{c}$.

The counterion condensation phenomenon can be only noticed at close
proximity of the charged cylinder, by the change of the short-distance
behavior of the potential from~(\ref{eq:y-short})
to~(\ref{eq:yshort-condens}). At large distances, the potential is
always given by~(\ref{eq:Xiren}), regardless if the counterion
condensation has taken place or not. When $\xi=\xi_{c}$, no
singularity appears in equation~(\ref{eq:Xiren}), nor in the
renormalized charge $\xi_{\ren}$ which characterizes only the
large-distance behavior of the potential. Also, notice that
$\xi_{\ren}\neq \xi_{\Manning}$ and $\xi_{\ren}\neq \xi_{c}$. This has
caused some confusion in the past, since in the original work
of~\cite{Manning}, in the condensed phase, the diffuse screening cloud
of the remaining uncondensed counterions around the charged cylinder
was treated using the linear Poisson--Boltzmann equation and using an
effective charge of the cylinder given by $\xi_{\Manning}=1$ to
account for the counterion condensation. It turns out that this
picture is not completely correct, since besides the counterion
condensation, there are also additional nonlinear effects in the
uncondensed cloud that are responsible for an additional charge
renormalization. As a result $\xi_{\ren}\neq \xi_{\Manning}$. For
example, in the limiting case $a=0$, from
equation~(\ref{eq:xiren-exact}), we obtain $\xi_{\ren}=2/\pi\simeq
0.63662 < \xi_{\Manning}=1$.

\acknowledgements

The author acknowledges partial financial support from Comit\'e de
Investigaciones y Posgrados, Facultad de Ciencias, Universidad de los
Andes.


\begin{thebibliography}
\bibitem[Alexander et al. 1984]{Alexander} Alexander, S., Chaikin,
  P.~M., Grant, P., Morales, G.~J. and Pincus, P. 1984 Charge
  renormalization, osmotic pressure, and bulk modulus of colloidal
  crystals: Theory. J. Chem. Phys. \textbf{80} 5776.

\bibitem[Bocquet et al. 2002]{Bocquet-JCP} Bocquet, L., Trizac, E. and
  Aubouy, M. 2002 Effective charge saturation in colloidal
  suspensions. J.~Chem.~Phys.~\textbf{117} 8138.

\bibitem[Chapman 1913]{Chapman} Chapman, D.~L. 1913 A contribution to
  the theory of electrocapillarity. Philos. Mag. \textbf{25} 475.

\bibitem[Debye and H\"uckel 1923]{DH} Debye, P. and H\"uckel E. 1923
  The theory of electrolytes. I. Lowering of freezing point and
     related phenomena. Phys. Z \textbf{24} 185. 

\bibitem[Derjaguin and Landau 1941]{DL} Derjaguin, B. and Landau,
  L. 1941 Theory of the stability of strongly charged lyophobic sols
  and of the adhesion of strongly charged particles in solutions of
  electrolytes. Acta Physico Chemica URSS \textbf{14}, 633.

\bibitem[Gouy 1910]{Gouy} Gouy, G.~L. 1910 Sur la constitution de la
  charge \'electrique a la surface d'un
  \'electrolyte. J. Phys.~\textbf{9}, 457.

\bibitem[Manning 1969]{Manning} Manning, G.~S. 1969 Limiting Laws and Counterion
    Condensation in Polyelectrolyte Solutions I. Colligative
    Properties. J.~Chem.~Phys.~\textbf{51} 924

\bibitem[McCaskill and Fackerell 1988]{McCaskill} McCaskill, J.~S.,
  and Fackerell, E.~D. 1988 Painlev\'e solution of the
  Poisson--Boltzmann equation for a cylindrical polyelectrolyte in
  excess salt solution. J.~Chem.~Soc.~Faraday Trans. 2 \textbf{84} 161

\bibitem[McCoy, Tracy and Wu 1977]{McCoy} McCoy, B.~M., Tracy, C.~A.,
  and Wu, T.~T. 1977 Painlev\'e functions of the third
  kind. J.~Math.~Phys.~\textbf{18} 1058.

\bibitem[Oosawa 1971]{Oosawa} Oosawa, F. 1971 \textit{Polyelectrolytes}.
  Dekker, New York.

\bibitem[Overbeek and Verwey 1948]{VO} Verwey, E.~J.~W. and
  Overbeek, J.~T.~G. 1948 \textit{Theory of the stability of
    lyophobic colloids}. Elsevier.

\bibitem[Shkel et al. 2000]{Shkel} Shkel, I.~A., Tsodikov, O.~V. and
  Record, M.~T. 2000 Complete asymptotic solution of cylindrical and
  spherical Poisson--Boltzmann equations at experimental salt
  concentrations. J. Phys. Chem. B \textbf{104} 5161.

\bibitem[T\'ellez and Trizac 2003]{Tellez-sat} T\'ellez, G., and Trizac, E. 2003
  Density functional theory study of the electric potential
  saturation: Planar geometry. Phys. Rev. E \textbf{68} 061401.

\bibitem[T\'ellez and Trizac 2004]{Tellez21} T\'ellez, G. and Trizac,
  E. 2004 Non-linear screening of spherical and cylindrical colloids: the
  case of 1:2 and 2:1 electrolytes. Phys. Rev. E \textbf{70} 11404.

\bibitem[T\'ellez and Trizac 2006]{TellezTrizac-PBcyl-exact} T\'ellez,
  G., and Trizac, E. 2006 Exact asymptotic expansions for the
  cylindrical Poisson--Boltzmann equation. J.~Stat.~Mech.~P06018.

\bibitem[Tracy and Widom 1997]{TracyWidom-poly} Tracy, C.~A., and Widom,
  H. 1997 On exact solutions to the cylindrical Poisson--Boltzmann
  equation with applications to polyelectrolytes. Physica A
  \textbf{244} 402.

\bibitem[Tracy and Widom 1998]{TracyWidom} Tracy, C.~A., and Widom,
  H. 1998 Asymptotics of a Class of Solutions to the Cylindrical Toda
  Equations. Comm.~Math.~Phys.~\textbf{190} 697.

\bibitem[Trizac et al. 2002]{TrizacPRL} Trizac, E., Bocquet, L. and
  Aubouy, M. 2002 A simple approach for charge renormalization of
  highly charged macro-ions. Phys. Rev. Lett. \textbf{89} 248301.

\bibitem[Trizac et al. 2003]{Trizac-analytical} Trizac, E., Aubouy,
  M., and Bocquet, L. 2003 Analytical estimate of effective charges at
  saturation in Poisson--Boltzmann cell models. J.~Phys.:
  Condens. Matt. \textbf{15} S291. 

\bibitem[Trizac and T\'ellez 2006]{TrizacTellez-Manning} Trizac, E.,
  and T\'ellez, G. 2006 Onsager-Manning-Oosawa Condensation Phenomenon
  and the Effect of Salt. Phys.~Rev.~Lett.~\textbf{96} 038302

\bibitem[Trizac and T\'ellez 2007]{TrizacTellez-macromol} Trizac,
  E. and T\'ellez, G. 2007 Preferential interaction coefficient for
  nucleic acids and other cylindrical polyions. Macromolecules \textbf{40}(4)
  1305.

\bibitem[Widom 1997]{Widom} Widom, H. 1997 Some Classes of Solutions to
  the Toda Lattice Hierarchy. Comm.~Math.~Phys.~\textbf{184} 653. 

\end{thebibliography}
\end{document}